\documentclass{article}
\usepackage{spconf,amsmath,graphicx,hyperref,framtex,xcolor}
\usepackage{cite}
\usepackage[normalem]{ulem}
\robustify\bfseries
\robustify\uline

\def\x{{\bm{x}}}

\def\eps{{\bm{\varepsilon}}}

\title{Virtual Consistency for Audio Editing}
\name{Matthieu Cervera$^1$, Francesco Paissan$\,^{1,2}$, Mirco Ravanelli$\,^{3,4,1}$, Cem Subakan$\,^{1,3,2}$\thanks{This research was enabled in part by support provided by the Digital Research Alliance of Canada (alliancecan.ca).}}
\address{ $^{1}$ Mila-Qu\'ebec AI Institute, $^{2}$ Laval University, $^{3}$ Concordia University, $^{4}$ University of Montreal}
\begin{document}
\ninept
\maketitle

\begin{abstract}

Free-form, text-based audio editing remains a persistent challenge, despite progress in inversion-based neural methods. Current approaches rely on slow inversion procedures, limiting their practicality. We present a virtual-consistency based audio editing system that bypasses inversion by adapting the sampling process of diffusion models. Our pipeline is model-agnostic, requiring no fine-tuning or architectural changes, and achieves substantial speed-ups over recent neural editing baselines. Crucially, it achieves this efficiency without compromising quality, as demonstrated by quantitative benchmarks and a user study involving 16 participants.

\end{abstract}


\section{Introduction}\label{sec:intro}

Neural Audio Editing (NAE) has recently gained increasing attention, driven by advances in generative modelling, with Diffusion Models (DMs) playing a central role ~\cite{liu2024visual, Liu2023AudioLDM2L}. 
While text-to-audio generation has reached remarkable levels of quality, with models such as {AudioLDM}~\cite{liu2023audioldm,Liu2023AudioLDM2L}, {AudioGen}~\cite{kreuk2022audiogen}, {MusicGen}~\cite{copet2024simple}, {MelodyFlow}~\cite{lan2024highfidelitytextguidedmusic} and {TANGO}~\cite{ghosal2023tango} producing convincing audio directly from prompts, many practical scenarios require more than generation from scratch. 
In real-world scenarios, users are typically interested in \emph{editing} existing recordings rather than generating entirely new ones. Such modifications require fine-grained control, enabling adjustments to content, removal of unwanted elements, or the addition of new sounds while preserving the integrity of the source material. Consequently, NAE has emerged as a crucial step beyond pure generative modelling.

Recent efforts have adapted latent diffusion models to provide editing capabilities. 
For example, {AUDIT}~\cite{wang2023audit} and InstructME~\cite{han2023instructmeinstructionguidedmusic} allow users to manipulate audio through a predefined set of operations, such as ``Add'' a sound, ``Drop'' a segment, or ``Replace'' excerpts. 
While effective in constrained scenarios, this design limits edits to a fixed set of commands, motivating research into more flexible frameworks. 
In \cite{paissan2024b_inspeech}, the authors adapt Imagic~\cite{kawar2023imagic} for audio by (i) solving an optimization problem to identify a feasible prompt reproducing the input audio, (ii) fine-tuning the diffusion model for exact reconstruction, and (iii) interpolating between the optimized prompt and the target edit. 
DITTO~\cite{novack2024ditto2distilleddiffusioninferencetime} optimizes the initial latent noise fed to a diffusion model by minimizing a loss between a target feature and the feature extracted from the sampled output. It enables selecting which main features of an audio should be preserved while editing, but it is restricted to differentiable extractors. 
SteerMusic~\cite{niu2025steermusic} adapts the delta denoising score method to guide a score-based model toward a target prompt or a personalized concept.
Although these methods extend the editing capabilities to non-rigid text prompts, they require substantial computational resources due to their multi-step optimization pipelines.

Another line of research has focused on \textit{inversion-based editing}, exemplified by SDEdit~\cite{Meng2021SDEditGI} and ZETA~\cite{Manor2024ZeroShotUA}. 
SDEdit applies iterative noising and denoising steps conditioned on the text prompt, avoiding the command-based limits of AUDIT. 
However, depending on the noise level, SDEdit may generate outputs that significantly diverge from the original input, limiting its fidelity to the input audio. 
ZETA, instead, employs the edit-friendly DDPM inversion paradigm to recover noise vectors associated with the input trajectory, and subsequently steers sampling toward the desired edit. 
To mitigate the reconstruction error caused by inversion, different works combined it with attention map manipulations, as shown by MEDIC~\cite{liu2024mediczeroshotmusicediting} andPPAE~\cite{xu2024promptguidedpreciseaudioediting}. One can observe that AP-Adapter~\cite{tsai2024audiopromptadapterunleashing} and MusicMagus~\cite{zhang2024musicmaguszeroshottexttomusicediting} also leverage attention map manipulations, but without inversion.
MelodyFlow~\cite{lan2024highfidelitytextguidedmusic} shares conceptual similarity with inversion methods and introduces an inversion mechanism compatible with an architecture trained on a flow-matching objective.

In this paper, we investigate how to achieve high-fidelity, non-rigid text-prompted edits for audio recordings. Our approach builds on an enhanced version of virtual consistency~\cite{infedit}, detailed in Section~\ref{sec:method}. Given an audio input, the core principle is to adapt the reverse sampling process of a diffusion model by reformulating it as consistency sampling. At each step, we compute, without parametrization, a noise vector that enables perfect reconstruction of the input during sampling. This formulation enables the computation of a closed-form solution to the inversion problem, yielding an editing pipeline that surpasses state-of-the-art methods across both qualitative and quantitative evaluation metrics with a fraction of the computational cost.

\noindent Our contributions are as follows:
\begin{itemize}
    \setlength\itemsep{.003cm}
    \item We show that inversion-free virtual consistency can be successfully used for neural audio editing, in a very computationally efficient way;
    \item We introduce a hyperparameter to control the edit strength in virtual consistency-based pipelines;
    \item We compare several recent audio editing models, and show that our approach outperforms the competing methods quantitatively and qualitatively. %
\end{itemize}

The paper structure is as follows: \cref{sec:method} presents our inversion-free editing pipeline, \cref{sec:experiments} presents the experimental setup, including the metrics and qualitative evaluation. Finally, in \cref{sec:results}, we discuss our findings.

\section{Inversion-free Audio Editing}\label{sec:method}

\subsection{Diffusion and Consistency Models}
Diffusion Denoising Probabilistic Models (DDPM)~\cite{ho2020denoising}, also referred to as diffusion models, consist of two processes:~the forward and the reverse (or generative) processes. During training, the DDPM gradually adds Gaussian noise to an initial data point, sampled from the data distribution $\x_0\sim \pp_\text{data}(\mathcal{X})$. This process is known as the forward process and is analytically defined by the transition
\begin{align}\label{eq:forward}
    \x_t &= \sqrt{\alpha_t}\,\x_{t-1} + \sqrt{\beta_t}\,\eps_t,
\end{align}
where $t \in \bracks{T}$ represents the timesteps from $1$ to $T$, the variance schedule is specified by $\beta_t\in(0,1)$, with $\alpha_t\defeq 1-\beta_t$, and $\eps_t\sim \mathcal{N}(0,\mathbf I)$ are i.i.d.\ noise vectors. Iterating the forward process gives the closed form
\begin{align}
    \x_t &= \sqrt{\bar\alpha_t}\,\x_0 + \sqrt{1-\bar\alpha_t}\,\epsilon, \qquad 
    \bar\alpha_t \defeq \prod_{s=1}^t \alpha_s,
\end{align}
which shows that $\x_T$ approaches an isotropic Gaussian distribution for sufficiently small $\bar\alpha_T$.  

During the reverse (or generative) process, a neural network ${ \eps_\theta: \mathcal{X}\times\bracks{T}\times\mathcal{C}\to\mathcal{X} }$ is trained to predict the noise injected at step $t\in\bracks{T}$ to the noisy sample $\x_t\in\mathcal{X}$, conditioned on the text prompt $c\in\mathcal{C}$. From the predicted noise, one can form an estimate of the clean input,
\begin{align}\label{eq:x0hat}
    \hat \x_0(\x_t,t) \;=\; \frac{\x_t - \sqrt{1-\bar\alpha_t}\,\eps_\theta(\x_t,t,c)}{\sqrt{\bar\alpha_t}}.
\end{align}
The true posterior $q(\x_{t-1}\mid \x_t,\x_0)$ is Gaussian with mean $\tilde\mu_t(\x_t,\x_0)$ and variance $\tilde\beta_t$, and substituting $\hat \x_0$ gives the parameterized reverse distribution
\begin{align}
    p_\theta(\x_{t-1}\mid \x_t) &= \mathcal{N}\parens*{\x_{t-1};\,\mu_\theta(\x_t,t),\,\sigma_t^2\mathbf I},\\
    \mu_\theta(\x_t,t) &= \frac{1}{\sqrt{\alpha_t}}\!\parens*{\x_t - \frac{\beta_t}{\sqrt{1-\bar\alpha_t}}\,\eps_\theta(\x_t,t,c)}.
\end{align}

Putting everything together, the general denoising update rule can be written as~\cite{ddim}
\begin{align}
    \x_{t-1} &= \sqrt{\bar\alpha_{t-1}}\,\frac{\x_t - \sqrt{1-\bar\alpha_t}\,\eps_\theta(\x_t,t,c)}{\sqrt{\bar\alpha_t}} \nonumber \\
    &\quad + \sqrt{1-\bar\alpha_{t-1}-\sigma_t^2}\:\eps_\theta(\x_t,t,c) + \sigma_t\,\eps_t,
    \label{eq:denoising}
\end{align}
where $\eps_t\sim\mathcal{N}(0,\mathbf I)$. This formulation separates three distinct components: the predicted clean sample (first term), a directional component toward the current noisy state (second term), and stochastic noise (third term). Setting $\sigma_t^2=\tilde\beta_t$ recovers the stochastic DDPM sampler, while $\sigma_t=0$ yields a deterministic trajectory, as explored in DDIM~\cite{ddim}.

More recently, Consistency Models (CM)~\cite{consistencymodels} were introduced as a way to trade off compute efficiency against sample quality. The key observation is that in diffusion models, the denoiser implicitly defines a mapping from a noisy sample $\x_t$ to a reconstruction of the clean input $\hat \x_0(\x_t,t)$. CMs make this connection explicit by directly learning a \emph{consistency function} ${f_\theta(\x_t,t)\approx \x_0, \: \forall \, t\in(\epsilon,T]}$, that should satisfy the self-consistency property defined as ${f_\theta(\x_t,t) = f_\theta(\x_{t'},t'), \: \forall\, t,t'\in(\epsilon,T]}$. This ensures that all denoised estimates along the trajectory agree on the same underlying $\x_0$. The boundary condition $f_\theta(\x_\epsilon,\epsilon)=\x_\epsilon$ ensures that very lightly noised inputs are mapped to themselves.  

With this formulation, CMs unify one-step and multi-step generation. In the one-step regime, a clean sample is obtained by evaluating $f_\theta(\x_T,T)$ for $\x_T\sim\mathcal{N}(0,\mathbf I)$. In the multi-step regime, one can alternately inject noise and apply the consistency function, yielding a trajectory that interpolates between the efficiency of direct mapping and the fidelity of iterative diffusion sampling.

\subsection{Virtual Consistency Inversion}\label{ssec:vci_edit}

The principle underlying inversion-based editing pipelines is that, instead of starting the generative process from pure noise, we first identify the forward trajectory $\{\x_t\}_{t=1}^T$ associated with an observed input $\x_0$, and then steer the reverse process towards the target edit. Concretely, the input $\x_0$ is inverted by applying the forward diffusion recursion \cref{eq:forward} so that we obtain a noisy sequence $(\x_1,\dots,\x_T)$ consistent with $\x_0$. During the reverse process, these noisy latents are then denoised while conditioning on the new editing prompt $c$.

To avoid the computational burden of explicitly performing such inversion, InfEdit~\cite{infedit} proposed a modification of the reverse dynamics. Specifically, in InfEdit, the authors set the sampling variance to $\sigma_t^2 = 1-\bar\alpha_{t-1}$, which eliminates the residual term (second term of the RHS) from the standard reverse update~\cref{eq:denoising}. The resulting update rule becomes
\begin{align}
    \x_{t-1} \;=\; \sqrt{\bar\alpha_{t-1}}\,f(\x_t,t;\x_0) \;+\; \sqrt{1-\bar\alpha_{t-1}}\,\epsilon_t, 
\end{align}
where the function $f$ plays the role of a consistency function:
\begin{align}\label{eq:cons-fn}
    f(\x_t,t;\x_0) \;=\; \frac{\x_t - \sqrt{1-\bar\alpha_t}\,\eps_\theta(\x_t,t,c)}{\sqrt{\bar\alpha_t}}.
\end{align}

Here, the self-consistency condition $f(\x_t,t;\x_0)=\x_0$ is enforced at each step without the need for the trained noise predictor $\eps_\theta$, by deriving the zero of \cref{eq:cons-fn}:
\begin{align}
    \eps_t^{\mathrm{cons}}(\x_t, \x_0) \;=\; \frac{\x_t - \sqrt{\bar\alpha_t}\,\x_0}{\sqrt{1-\bar\alpha_t}}.
\end{align}
By construction, this noise estimate guarantees perfect reconstruction of the original sample $\x_0$, making the method suitable for editing but not for unconditional generation. Due to its inversion-free nature, this approach is referred to as \emph{Virtual Consistency Inversion} (VCI)~\cite{infedit}.

To perform the edit, at every step $t$ we steer $\eps_t^{\mathrm{cons}}$ to guide the sampling process towards the desired edit. Concretely, we initialize the trajectory from a noisy input $\x_T^{\mathrm{src}} = \x_T^{\mathrm{tgt}} \sim \mathcal{N}(\mathbf{0}, \mathbf{I})$ and, given embeddings of a text prompt describing the source audio $c_{\mathrm{src}}\in\mathcal{C}$ and a prompt explaining the edit $c_{\mathrm{tgt}}\in\mathcal{C}$, we estimate the noise of the source and the target branches via $\eps_\theta$:
\[
\eps_t^{\mathrm{src}} = \eps_\theta(\x_t^{\mathrm{src}}, t, c_{\mathrm{src}}), \qquad
\eps_t^{\mathrm{tgt}} = \eps_\theta(\x_t^{\mathrm{tgt}}, t, c_{\mathrm{tgt}}).
\]
Intuitively, the difference $\Delta\eps_t \coloneqq \eps^{\mathrm{tgt}}_t - \eps^{\mathrm{src}}_t$ contains the information regarding the changes required to $\eps_t^{\mathrm{src}}$ to represent the target edit. Therefore, to perform the edit, we update $\x_t^{\mathrm{tgt}}$ by performing one step of the reverse process using $\eps_t^{\mathrm{edit}}$ defined as
\begin{align}\label{eq:controlvci}
    \eps_t^{\mathrm{edit}}\defeq\frac{\varphi}{\sqrt{2}}\Delta\eps_t+\sqrt{1-\varphi^2}\:\eps_t^{\mathrm{cons}}(\x_t, \x^{\mathrm{src}}_t),
\end{align}
where $\varphi\in\bracks{0,1}$ controls the edit strength and $\x_t^{\mathrm{src}}$ is explicitly denoised using the input $\x_0$ at each step. We emphasize that our formulation departs from the original VCI editing approach by introducing a hyperparameter $\varphi$ to balance $\Delta\eps_t$ and $\eps_t^\mathrm{cons}$ (i.e., edit strength), with the balance enforced via a variance-constrained parametrization. By matching the second moment, the edited noise remains within the high-density annulus of the Gaussian prior, allowing the network to process edits at familiar magnitudes and preventing the directionality from being unintentionally amplified or suppressed. We denote this formulation as ControlVCI, in contrast to the original VCI, which defines ${\eps_t^{\mathrm{edit}}=\Delta\eps_t+\eps_t^{\mathrm{cons}}}$ \cite{infedit}.

\section{Experiments}\label{sec:experiments}
\begin{table*}[t]
    \centering
    \renewcommand{\arraystretch}{1.2}
    \centering
    \resizebox{.9\linewidth}{!}{
        \begin{tabular}{l|SS|SSSS|S}
        \toprule
        \textbf{Method} & \textbf{MuLan $\uparrow$} & \textbf{CLAP $\uparrow$} & \textbf{LPAPS $\downarrow$} & \textbf{FAD $\downarrow$} & \textbf{CQT-PCC $\uparrow$} & \textbf{Audiobox-AE $\downarrow$} & \textbf{Latency [\si{\s}]} \\
        \midrule
        \multicolumn{8}{c}{ZoME Bench} \\
        Input Audio & \textbf{0.318} & \text{0.248} & \text{0.0}   & \text{0.0} & \text{1.0} & \text{0.0} & N/A \\
        DDIM $T_\text{start}=80$ & \text{0.258} & \text{0.285} & \text{4.248}  & \text{0.498} &  \textbf{0.497} & \text{2.465} & \text{16.164}\\
        SDEdit $T_\text{start}=50$ & \text{0.225} & \text{0.280} & \text{5.991} &  \text{0.711} & \text{0.218} & \text{4.174} & \text{\phantom{0}8.798} \\
        ZETA $T_\text{start}=70$ & \text{0.267} &  \text{0.305} &  \text{4.897} &  \text{0.672} &  \text{0.366} & \text{3.166} & \text{23.758} \\
        MusicGen & \text{0.267}  &  \textbf{0.335} & \text{6.548} & \text{0.615} & \text{0.024} & \text{4.036} & \text{\phantom{0}9.245} \\
        VCI (Ours) & \text{0.279}  & \text{0.305} & \underline{3.961} & \underline{0.476} & \text{0.466} & \underline{2.426} & \textbf{\phantom{0}1.615} \\
        ControlVCI (Ours) & \underline{0.283} & \underline{0.309} & \textbf{3.761} & \textbf{0.475} & \underline{0.471} & \textbf{1.902} & \phantom{0}\underline{1.631} \\ \midrule
        \multicolumn{8}{c}{MedleyDB} \\
        Input Audio & \text{0.166} & \text{0.148} & \text{0.0}   & \text{0.0} & \text{1.0} & \text{0.0} & N/A \\
        DDIM $T_\text{start}=100$  &  \text{0.260} & \text{0.250} &  \textbf{5.003} &  \textbf{1.146} & \textbf{0.445} & \textbf{2.870} & \text{43.481}\\
        SDEdit $T_\text{start}=90$ & \text{0.290} & \text{0.280} & \text{6.120} &  \text{1.343} &  \text{0.219} & \text{4.055} & \text{22.393} \\
        ZETA $T_\text{start}=80$ &  \text{0.284} & \text{0.278} & \text{5.378} & \text{1.231} &   \underline{0.356} & \underline{3.239} & \text{64.667} \\
        MusicGen & \text{0.238} & \text{0.238} & \text{6.299} & \underline{1.177} & \text{0.030} & \text{5.339} & \text{35.689} \\
        VCI (Ours) & \textbf{0.313} & \textbf{0.294} & \text{5.465} &  \text{1.210} & \text{0.293} & \text{3.333} & \underline{12.601} \\
        ControlVCI (Ours) & \underline{0.302} & \underline{0.291} &  \underline{5.311} &  \text{1.206} &  \text{0.293} & \text{3.621} & \textbf{12.483} \\
        \bottomrule
        \end{tabular}
    }
    \vspace{-3pt}
    \caption{\textbf{Quantitative Evaluation.} Comparison of text and audio alignment metrics for our editing pipelines and the reference methods. Bold numbers indicated the best value for that metric, while underlined numbers indicate the second-best. The row denoted as ``Input Audio" refers to the metrics computed on the unedited audio files. For ControlVCI, we choose $\varphi=0.61$, the guidance scale $w_\text{tgt}=15.0$ for ZoME Bench and $\varphi=0.82$ $w_\text{tgt}=20.0$ for MDB. For both VCI and ControlVCI, we used 8 steps for ZoME and 20 steps for MedleyDB.}
    \label{tab:results_zome}
    \vspace{-7pt}
\end{table*}

\subsection{Evaluation Metrics}

In Neural Audio Editing, we modify an input audio according to user instructions, often expressed as textual prompts. A major challenge is ensuring the output both preserves important aspects of the original audio and reflects the intended edits. Different edits emphasize these objectives to varying degrees: some primarily alter timbre or style while keeping the overall structure intact, while others change the content or temporal organization itself. This makes it essential to evaluate edits using metrics that capture both fidelity to the input and adherence to the requested transformation.

\noindent \textbf{Audio Fidelity Metrics.} Similarly to \cite{paissan2024b_inspeech}, we use LPAPS, a point-to-point distance that measures the semantic alignment of two audio samples. Specifically, we used the intermediate representations from the audio encoder in the CLAP model. Additionally, we follow \cite{hou2025editing,steermusic} and use CQT-PCC and extract the top 1 CQT bins that contain most of the melody information. We employ the Fr\'echet Audio Distance (FAD) to compute the distance between the sets of original and edited audio samples. Finally, we use Meta Audiobox-Aesthetics (MAA) \cite{tjandra2025meta}, a no-reference audio quality predictor, to evaluate the aesthetics fidelity of the edits. MAA returns a tuple representing scores for production quality, production complexity, content enjoyment, and content usefulness. To compute an aesthetics-fidelity metric, we average the absolute difference of the MAA scores computed on the edited audio and the input audio.

\noindent \textbf{Text Consistency Metrics.} As is now common practice in the neural editing literature, we used the CLAP~\cite{wu2024largescalecontrastivelanguageaudiopretraining} score between the requested edit prompt and the edited audio. Additionally, we use MuQ-MuLan \cite{zhu2025muq}, a joint music-text embedding model based on contrastive learning to compute the cosine similarity between the edited audio and the target text embedding. For CLAP, we used the \texttt{music\_audioset\_epoch\_15\_esc\_90.14} checkpoint, while for MuQ-MuLan, we used the \texttt{OpenMuQ/MuQ-MuLan-large} checkpoint.

\subsection{Datasets}
We used the ZoME Bench~\cite{liu2024mediczeroshotmusicediting} dataset, a subset of MusicCaps that contains 943 (audio, source prompt, edit prompt) tuples. Each audio file is \SI{10}{\s} long. The source prompts from MusicCaps are refactored by ChatGPT-4, while the target prompts are directly generated by ChatGPT-4. Additionally, following \cite{Manor2024ZeroShotUA}, we used the MusicDelta subset of MedleyDB~\cite{bittner2016medleydb}, ``MedleyMDPrompts". This contains 259 edits with the corresponding source and target edit prompts. These edits were generated from 13 audio samples between \SI{17}{\s} to \SI{2}{\min} long. The ground-truth labels and edit prompts were manually labelled by the authors of ZETA.

\subsection{Reference Methods}
We compare our pipeline with DDIM partial inversion, as we observed that the full inversion generated worse results, likely because of the accumulation of inversion error. Additionally, we used the official implementations of SDEDIT, ZETA, and MusicGen. For inversion-based methods, we set the guidance scales $w$ to match the values used by ZETA as they also seemed to yield the best results in our experiments ($w_\text{src}= 3.0$ for each method, $w_\text{tgt}^\text{ZETA}=w_\text{tgt}^\text{SDEDIT}=12.0$ and $w_\text{tgt}^\text{DDIM}=5.0$). However, we had to slightly adjust $T_\text{start}$ values. For MusicGen, we used the \texttt{facebook/musicgen-melody} checkpoint to condition on a text and on the input audio features, while for all the diffusion-based pipelines we used AudioLDMv2\cite{Liu2023AudioLDM2L} with 200 sampling steps. We report the inference speeds measured on ``3/8th of the computing power of an H100" with 40GB GPU memory\footnote{This requirement derives from the way allocations are handled on the FIR cluster \url{https://docs.alliancecan.ca/wiki/Fir}.}. Specifically, we measure the latency required to go from the input to the output of the autoencoder, and we average it on the whole dataset. Note that we do not compare our pipeline to SteerMusic, MelodyFlow, and MEDIC as they have not released their codebase.

\sisetup{
  round-mode           = places,
  round-precision      = 3,
  detect-weight        = true,
  table-text-alignment = center
}

\newcommand{\alignedunderline}[1]{%
  \makebox[0pt][c]{\underline{\makebox[\widthof{\num{#1}}][c]{\num{#1}}}}%
}

\subsection{Qualitative evaluation.}

To further ensure that the comparison is fair, we performed a user study to evaluate the perceived text and audio alignments. Concretely, we conducted a WebMUSHRA \cite{Schoeffler-2018} questionnaire, presenting users with nine edits randomly picked from the ZoME and MedleyDB benchmarks. Users were required to evaluate, on a scale from 0 to 100, the (i) alignment to the input audio and (ii) the alignment to the target text. We make the audio samples presented during the user study available on our companion website\footnote{\url{https://matthieu-cervera-9e056d.gitlab.io/vci_editing}}.

\section{Results}\label{sec:results}

\subsection{Quantitative Results}

\begin{figure}[t]
    \centering
    \resizebox{\linewidth}{!}{
        \includegraphics[width=0.49\linewidth]{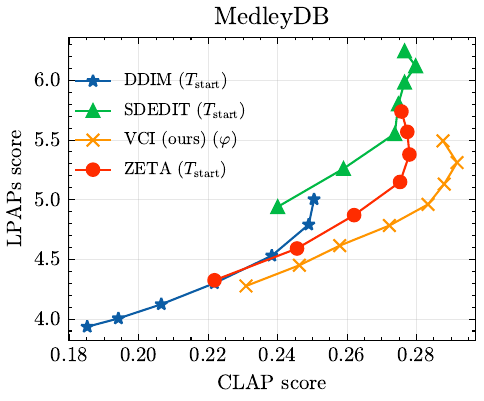}
        \includegraphics[width=0.49\linewidth]{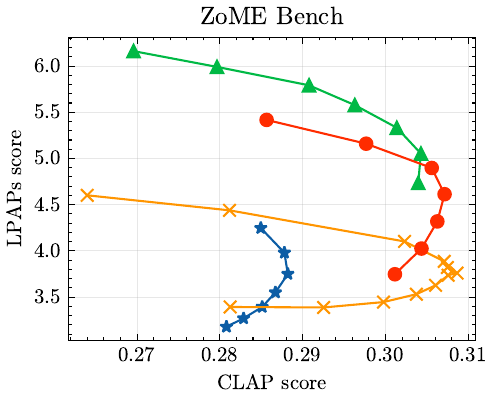}
    }
    \vspace{-.6cm}
    \caption{\textbf{Text-Audio Fidelity Trade-off.} Each dot represents a different edit strength. The dots are connected in order of increasing parameter values. The values of $T_\text{start}$ range from explored from 40 (left-most dot) to 100 (end of the line). For ours, we explore $\varphi\in\bracks{0.10, 0.95}$.}
    \label{fig:edit-strength}
    \vspace{-15pt}
\end{figure}

In \cref{tab:results_zome}, we summarize the results obtained on both the ZoME Bench and MedleyDB datasets. On ZoME Bench, we observe that ControlVCI outperforms all reference methods in terms of fidelity to the input audio, while VCI scores second. In text alignment, MusicGen obtains the best CLAP score, with VCI and ZETA following closely. The MuLan score, by contrast, performs best on unedited audio files, highlighting that editing benchmarks relying on AI-generated captions may not be as reliable as those based on human annotations. Following this observation, we decided to complement our results with a user perception analysis. Our findings on ZoME Bench indicate that ControlVCI achieves the most favourable balance between input fidelity and text fidelity, surpassing all reference methods when both criteria are considered jointly.

On the MedleyDB benchmark, we find that DDIM inversion provides the strongest alignment with the input audio. Nonetheless, we observe that DDIM's performance in text alignment is not good, suggesting that the model is not performing the desired edit. VCI and ControlVCI, on the other hand, strike a better balance. In fact, VCI achieves the best alignment with the text prompt, and ControlVCI the second-best. 
Also for MedleyDB, our method achieves the best trade-off between text and audio alignment.

Finally, it is worth emphasizing that VCI-based pipelines operate with substantially lower latency than alternative approaches, offering a compelling combination of efficiency and effectiveness. %

\subsection{Text-Audio Alignment Trade-off}

To further examine the interplay between text and audio fidelity, we perform a trade-off analysis in \cref{fig:edit-strength}, illustrating how varying the edit strength influences this balance. Methods that rely on changing the number of timesteps provide only a coarse mechanism for control: reducing timesteps often leads to under-editing, while increasing them can result in excessive distortion, making it difficult to precisely target intermediate regimes. In contrast, ControlVCI introduces the $\varphi$ parameter, which directly modulates the predicted noise and thus allows for fine-grained, continuous adjustments of the edit strength. This enables the model to more reliably navigate the trade-off space, consistently achieving a better balance between text fidelity and audio fidelity across both datasets.

\subsection{Qualitative Results}

\begin{figure}[t]
    \centering
    \includegraphics[width=\linewidth]{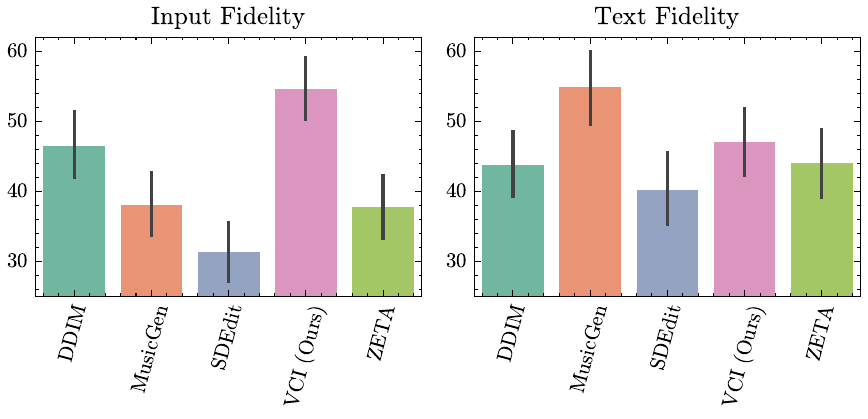}
    \vspace{-.6cm}
    \caption{\textbf{User Study Results.} From left to right, we present the perceived overall preference, input fidelity, and text fidelity.}
    \label{fig:user-study}
    \vspace{-15pt}
\end{figure}

\Cref{fig:user-study} presents the outcomes of our user study, conducted with 16 participants. The results indicate that our pipeline achieves the highest ratings in terms of input fidelity, while ranking second with respect to text fidelity. This suggests that the proposed approach excels at preserving speaker identity and acoustic consistency, even when modifications are introduced. By contrast, MusicGen scored higher in text fidelity but we observed that MusicGen suffers significantly in terms input fidelity, generating outputs that deviate significantly from the input. 
Importantly, when both evaluation dimensions are considered jointly, VCI emerges as the most effective pipeline overall. This balance highlights the value of our design choices: rather than optimizing a single dimension, VCI delivers robust performance across multiple user-relevant criteria. Such results underscore the suitability of our approach for practical audio editing scenarios, where both faithfulness to the original recording and accurate realization of the intended edits are equally critical.

\section{Conclusion}
In this paper, we introduced an inversion-free neural audio editing pipeline that enables text-guided modifications of audio signals. We demonstrate that our approach effectively balances the trade-off between input fidelity, \ie, the preservation of the original audio characteristics, and text fidelity, \ie., the alignment with the provided editing instructions. This balance is evaluated both quantitatively, through objective metrics, and qualitatively, via perceptual studies. Moreover, we show that our pipeline attains these improvements while requiring only a fraction of the computational resources compared to the reference methods, thereby making it more accessible and scalable for practical applications.

\vfill
\pagebreak

\bibliographystyle{IEEEbib-short}   %
\bibliography{strings}

\end{document}